\documentclass[aps, twocolumn, showpacs, preprintnumbers, superscriptaddress, amsmath, amssymb, prb]{revtex4}
\usepackage{graphicx}
\usepackage{dcolumn}
\usepackage{bm}
\usepackage{color}
\usepackage{amsfonts}

\begin{document}

\title{ On the Truncation Error of Numerical Renormalization Group}

\author{Ke Yang}
\affiliation{Department of Physics, Renmin University of China, 100872 Beijing, China}
\author{Ning-Hua Tong}
\email{nhtong@ruc.edu.cn}
\affiliation{Department of Physics, Renmin University of China, 100872 Beijing, China}
\date{\today}

\begin{abstract}
Using the recently developed exact numerical renormalization group (NRG) method, we analyse the NRG truncation errors $\delta \chi$ of the local magnetic susceptibility and $\delta F$ of the free energy for the spin-boson model (SBM). We find that for temperatures higher than a crossover temperature $T_{cr}$, as the number of kept states $M$ increases, both errors have oscillations with quasi period $\ln{M}/\ln{N_b}$ and the envelopes decrease as $\epsilon_{tr}=\Lambda^{-\ln{M}/\ln{N_b}}$ ($N_b$ is the number of boson states used for each bath site). For $T \ll T_{cr}$, they decrease slower than the power law. We extract that $T_{cr} = T^{\ast} \epsilon_{tr}$, with $T^{\ast}$ being the crossover energy scale between the declocalized and the critical fixed points of SBM. The same rule applies to $\delta \chi$ and $\delta F$ calculated from the full density matrix NRG method and is expected to hold for general impurity models, allowing accurate removal of NRG truncation errors in static quantities at high temperatures.
\end{abstract}

\pacs{05.10.Cc, 05.30.Jp, 71.27.+a}


\maketitle

\section{Introduction}

Wilson's numerical renormalization group (NRG) \cite{Wilson1,Bulla1} is the method of choice for studying quantum impurity models that describe small quantum systems subjected to the influence of continuous non-interacting environmental degrees of freedom (bath). Such quantum impurity systems appear in many contexts, including the Kondo physics,\cite{Kondo1} quantum dot,\cite{Jeong1} superconducting qubit,\cite{Makhlin1} photosynthetic biosystem,\cite{Muehlbacher1} and the dynamical mean-field theory,\cite{Georges1,Vollhardt1} etc.

NRG consists of three steps, the logarithmic discretization of the bath degrees of freedom, canonical transformation of Hamiltonian into the form of a semi-infinite chain, and iterative diagonalization. Two approximations are made in these steps. In the first step, the continuous bath degrees of freedom are approximately coarse grained into discrete ones with exponentially descending energy scales. The introduced logarithmic discretization error is controlled by $\Lambda \geqslant 1$ and decreases to zero as $\Lambda$ tends to unity. In the third step, the full eigenvalue spectrum is truncated to the lowest $M$ states after each diagonalization. It introduces the truncation error, which is controlled by the number of kept states $M$ and diminishes as $M$ tends to infinity. Numerically exact results are obtainable only in the simultaneous limit $\Lambda=1$ and $M=\infty$. In most situations, these two errors influence the NRG results only at quantitative level and qualitative conclusions from NRG are reliable. But exceptions do exist.\cite{Vojta1} For a reliable and accurate study of quantum impurity systems, it is important to check the conclusion by extrapolating NRG data to the exact limit.

Since the logarithmic discretization error is introduced at the stage of logarithmic discretization, it only depends on $\Lambda$. It can be reduced by modified discretization methods\cite{Campo1,Zitko1} or by the $z$-averaging scheme.\cite{Yoshida1} In contrast, the truncation error seems more difficult to handle: it depends not only on $M$ but also on $\Lambda$. It is due to the principle of NRG that the truncation error increases as $\Lambda$ decreases for a fixed $M$. This hinders the extrapolation of NRG results to the simultaneous limit of $\Lambda=1$ and $M=\infty$. Systematic and universal analysis of the truncation error is still lacking.

In this work, we study the NRG truncation errors $\delta \chi$ of the local magnetic susceptibility $\chi$ and $\delta F$ of the free energy $F$ for the spin-boson model (SBM).\cite{Leggett1} By detailed analysis of their dependence on $M$, $\Lambda$, and the truncated Hilbert space dimension $N_b$ of the boson bath site, we find a quantitative description for these truncation errors. For temperatures higher than a crossover temperature $T_{cr}$, both $\delta \chi$ and $\delta F$ oscillate with the quasi period $\ln{M}/\ln{N_b}$. The envelopes decrease as $\epsilon_{tr} = \Lambda^{-\ln{M}/\ln{N_b}}$ with increasing $M$. For $T \ll T_{cr}$, they decrease slower than the power law. The crossover temperature is found to be $T_{cr} = T^{\ast} \epsilon_{tr}$, with $T^{\ast}$ being the crossover energy scale between the declocalized and the critical fixed points of SBM. We draw this conclusion from the data produced by the newly developed exact NRG (ENRG) method,\cite{Yang1} which avoids additional approximations in calculating physical quantities from the NRG eigen values and eigen states. We have checked that this conclusion also holds for the full density matrix (FDM) NRG method.\cite{Weichselbaum1,Peters1} With a clear physical picture behind this result, we expect that the conclusion applies to general quantum impurity models. Our result makes it possible to accurately remove the NRG truncation errors in static quantities at high temperatures.

\section{Model and Method}

For studying the NRG truncation errors, we use the SBM which is a generic quantum impurity model. SBM describes a two-level quantum system subjected to the influence of a dissipative bosonic bath.\cite{Leggett1,Weiss1} It has been widely studied in various fields ranging from the superconducting qubit system\cite{Makhlin1} to photosynthetic biosystems. \cite{Muehlbacher1} NRG has played an important role in the understanding of this model.\cite{Bulla2,Vojta2,Guo1,Tong1} The Hamiltonian of SBM reads 
\begin{equation}
   H_{SB} = -\frac{\Delta}{2} \sigma_x + \frac{\epsilon}{2}\sigma_z + \sum_i \omega_i a_i^{\dagger}a_i + \frac{\sigma_z}{2}\sum_{i}\lambda_i \left(a_i^{\dagger}+a_i \right). 
\end{equation}
The two-level system is described by Pauli matrices and the influence of bath is encoded into the spectral function $J(\omega) = \pi \sum_i \lambda_i^{2} \delta(\omega - \omega_i)$, for which we use 
\begin{equation}
J(\omega) = 2 \pi \alpha \omega^{s} \omega_c^{1-s}  \,\,\,\,\,\, (0 < \omega < \omega_c).
\end{equation}
$\omega_c=1.0$ is set as the energy unit. The coupling strength between the spin and the bath is described by $\alpha$. $s$ is the exponent describing the density of low energy bath modes. 
The free energy $F$ that we will study is defined as $F=-(1/\beta)\ln{Z}$, with the partition function $Z= Tr\left( e^{-\beta H_{SB}} \right)$.
The local magnetic susceptibility is defined as\cite{Note1}
\begin{eqnarray}
 \chi & \equiv & -\frac{1}{2}\frac{ \partial \langle \sigma_z \rangle }{ \partial \epsilon}|_{\epsilon=0} \nonumber \\
 &=& -\frac{1}{4} \text{Re} G^{b}_{\sigma_z \sigma_z}(\omega =0) - \frac{\beta}{4} \langle \sigma_z \rangle^2.
\end{eqnarray}
In this equation, $G^{b}_{\sigma_z \sigma_z}(\omega)$ is the retarded Bose-type Green's function (GF) of operator $\sigma_z$. For general operators $A$ and $B$, the Bose-type retarded GF is defined as
\begin{equation}
G^{b}_{A,B}(\omega) \equiv -i \int_{0}^{+\infty} \langle \left[ A(t), B(t^{\prime}) \right]\rangle e^{i(\omega+ i\eta)(t-t^{\prime})} d(t-t^{\prime}).
\end{equation}
The bracket $\left[X,Y \right]$ in Eq.(4) denotes the commutator of $X$ and $Y$.

For this model, the Hamiltonian of the semi-infinite chain (truncated to length $N$) obtained through the canonical transformation reads
\begin{eqnarray}
   H_{N} &=& -\frac{\Delta}{2} \sigma_x +  \frac{\epsilon}{2}\sigma_z 
   + \frac{1}{2}\sqrt{\frac{\eta_0}{\pi}} \sigma_z \left( b_0 + b_0^{\dagger} \right)  \nonumber \\
   &+& \sum_{n=0}^{N} \epsilon_n b_{n}^{\dagger}b_n + \sum_{n=0}^{N-1} t_n \left( b_n^{\dagger} b_{n+1} + b_{n+1}^{\dagger} b_n \right).
\end{eqnarray}
Here, the parameters $\eta_0$, $\epsilon_n \propto \Lambda^{-n}$ and $t_n \propto \Lambda^{-n}$ are expressed in terms of $J(\omega)$ by standard formula.\cite{Bulla2} As usual, we truncate the Hilbert space of each boson site to $N_b$ states in the occupation basis.\cite{Bulla2} The NRG truncation errors are now functions of $\Lambda$, $M$, and $N_b$. 

To calculate the local magnetic susceptibility $\chi$ and the free energy $F$ using NRG, we have to express them in terms of the eigen states and eigen energies generated by the standard iterative diagonalization of NRG. For $\chi$, there are methods at various sophistication for this purpose. In the patching method,\cite{Bulla3} contributions from every energy shell are patched up by an approximate way and the sum rule of the spectral function is not guaranteed. In the density-matrix NRG method developed by Hofstetter,\cite{Hofstetter1} the reduced density matrices of the full system are used as the weights for the patching to fully take into account the influence of low energy states to high frequency spectral function. In the state of art FDM NRG method,\cite{Weichselbaum1,Peters1,Weichselbaum2} the Lehmann representation of spectral function is treated with the complete set of eigenstates of NRG\cite{Anders1} and simplified using the NRG approximation.\cite{Weichselbaum1} This method fulfils the sum rule of spectral function rigorously, but may produce negative diagonal spectral function in certain situations.\cite{Yang1} 
Less attention has been paid to the calculation of the free energy $F$ of the full system. FDM NRG provides a formula for $F$ based on the complete basis set of eigenstates composed of the discarded states in each NRG iteration.

In this work, we use the recently developed ENRG method\cite{Yang1} to calculate $\chi$ and $F$. ENRG method is based on the fact that the iterative diagonalization algorithm of NRG defines an exactly solvable Hamiltonian $\tilde{H}_{N}$, dubbed NRG Hamiltonian, whose eigen states and eigen energies can be obtained exactly from the iterative diagonalization process of NRG. 
The NRG Hamiltonian for a given chain length $N$ reads
\begin{eqnarray}
  \tilde{H}_{N} &=& H_{n_{0}} + \sum_{n=n_{0}+1}^{N} \epsilon_{n} b_{n}^{\dagger} b_{n}  \\
    && + \sum_{n=n_{0}}^{N-1}  t_n  \left[ \left(P_n b_{n}^{\dagger}P_n \right) b_{n+1} + b_{n+1}^{\dagger} \left(P_n b_{n} P_n \right) \right]. \nonumber 
\end{eqnarray}
Here, $P_n$ projects any state in the Hilbert space of $\tilde{H}_n$ into the subspace spanned by the kept eigen states of $\tilde{H}_n$. $H_{n_0}$ is the NRG Hamiltonian of the longest chain whose all eigen states are kept in the NRG calculation.
ENRG produces the exact physical quantities of $\tilde{H}_{N}$. The logarithmic discretization error and the truncation error in the ENRG results are inherited from those in $\tilde{H}_{N}(\Lambda, M, N_b)$. They are well-controlled by the NRG parameters $\Lambda$, $M$, and $N_b$, making it possible that they could be analysed and understood. Such analysis  provides the basis for reliable extrapolation of NRG results to the exact limit $\Lambda=1$ and $M=\infty$.\cite{Zitko2}
For $\chi$, the ENRG formula is different from that of FDM NRG and is expected to be more systematic in the $M$ dependence. For $F$, the ENRG formula is identical to that of FDM NRG. 
Details of ENRG method is presented in Ref.\onlinecite{Yang1} and we refer the interested readers to that paper. In this paper, after presenting the main results of ENRG, we compare $\delta \chi$ obtained from ENRG and FDM NRG and show that our conclusion applies to the FDM NRG method as well.

\section{Truncation Error At $\alpha = 0$} 

The truncation error $\delta \chi = \chi(M) - \chi(\infty)$ is best demonstrated at $\alpha=0$, where the spin is decoupled from the bath in $H_{N}$ but is coupled to it in $\tilde{H}_{N}$ at finite $M$ due to the spectrum truncation. At $\alpha=0$, $\chi(\infty)=\chi_{\text{exact}}=\tanh{(\beta \Delta/2)}/(2\Delta)$ does not depend on $\Lambda$ and $N_b$. $F(M=\infty)$ depends on $\Lambda$ and $N_b$ due to the bath contribution.

\begin{figure}[t!]
\vspace{-2.0cm}
\begin{center}
\includegraphics[width=5.0in, height=3.3in, angle=0]{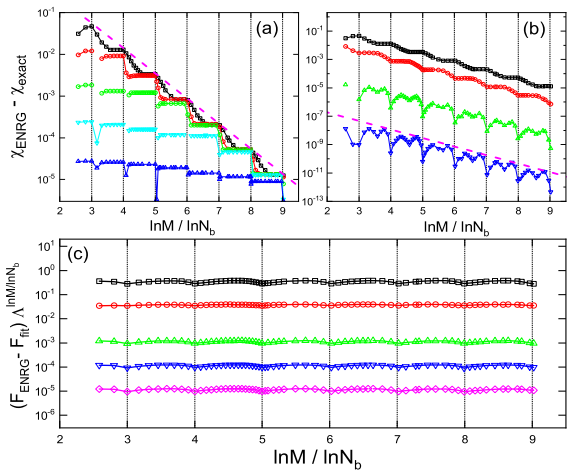}
\vspace*{-1.0cm}
\end{center}
\caption{ENRG results for $\delta \chi$ (in (a) and (b)) and $\delta F\Lambda^{\ln{M}/\ln{N_b}} $ (in (c)) at $\alpha=0$, as functions of $\ln{M}/\ln{N_b}$ for different temperatures. From top to bottom, (a): $T/\Delta=10^{-1}$, $10^{-2}$, $10^{-3}$, $10^{-4}$, and $10^{-5}$; (b): $T/\Delta=10^{-1}$, $10^{0}$, $10^{1}$, and $10^{2}$; (c): $T/\Delta=10^{1}$, $10^{0}$, $10^{-1}$, $10^{-2}$, and $10^{-3}$. The dashed lines are $y=3.6\Lambda^{-x}$ in (a) and $y=3.0 \times 10^{-6} \Lambda^{-x}$ in (b). In (c), $F_{\text{fit}}$ is obtained by fitting Eqs.(7)-(8). Other parameters are $s=0.3$, $\Delta=0.08$, $\epsilon=0.0$, $\Lambda=4.0$, $N_b=2$, and $N=16$. }
\label{Fig1}
\end{figure}

%
\begin{figure}[t!]
\vspace*{-3.5cm}
\begin{center}
\includegraphics[width=6.2in, height=4.5in, angle=0]{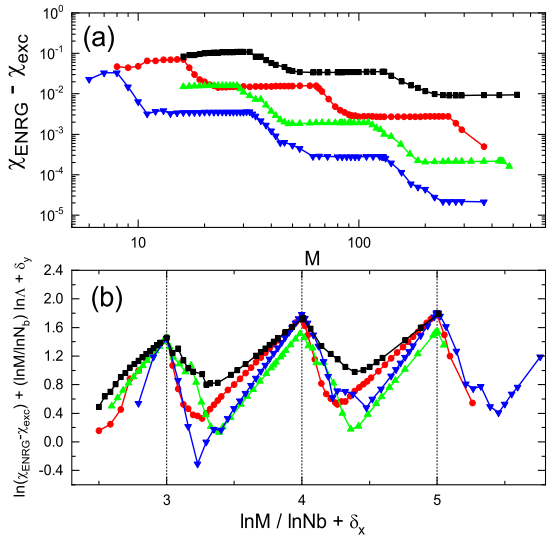}
\vspace*{-1.0cm}
\end{center}
\caption{ (a) Truncation error in the local magnetic susceptibility $\chi$ as functions of $M$ for various $\Lambda$ values at $\alpha=0.0$, produced by ENRG. (b) Replot of (a) on rescaled $x$ and $y$ axsis. $\delta_x$ and $\delta_y$ are chosen such that the data point of all curves coincide at $\ln{M}/\ln{N_b}+\delta_x=3.0$. $\Lambda$ values are $4.0$ (black squares); $6.0$ (red circles); $9.0$ (green up triangles), and $13.0$ (blue down triangles). Lines are for guiding eyes. Other parameters are $s=0.3$, $\Delta=0.08$, $\epsilon=0.0$, $\alpha=0.0$, $T=0.1\Delta$, and $N_b =4$.}   \label{Fig2}
\end{figure}

As shown in Fig.1(a) (low $T$) and 1(b) (high $T$) for $N_b=2$, abrupt changes occur in the curve $\delta \chi(M)$ at integer values of $\ln{M}/\ln{N_b}$, i.e., at $M = N_b^{k}$ ($k=1,2,...$). The envelope of the curve has a temperature-dependent behavior. At high $T$ (in Fig.1(b)), apart from a decaying power-law factor $\Lambda^{-\ln{M}/\ln{N_b}}$ (supported further by data shown in Fig.2 and Fig.3), $\delta \chi$ is quasi periodic in $\ln{M}/\ln{N_b}$, with period unity. At low $T$ (in Fig.1(a)), the envelope curve has a crossover from the slow decay in small $M$ to the behavior of $\Lambda^{-\ln{M}/\ln{N_b}}$ in large $M$ regime. The behavior can be summarized for $T \gg T_{cr}$ as
\begin{equation}
  \delta X (\Lambda, M, N_b) = f(\Lambda, \frac{\ln{M}}{\ln{N_b}}, N_b) \epsilon_{tr}(\Lambda, M, N_b).
\end{equation}
Here $X=\chi$. The periodicity $f(\Lambda, x, N_b) = f(\Lambda, x+1, N_b)$ does not hold strictly but asymptotically in the large $M$ limit. The dimensionless factor $\epsilon_{tr}(\Lambda, M, N_b)$ reads
\begin{equation}
\epsilon_{tr}(\Lambda, M, N_b) =\Lambda^{-\frac{\ln{M}}{\ln{N_b}}}.
\end{equation}
$T_{cr}$ is an $M$-dependent crossover temperature. For low temperature $T \ll T_{cr}$, the enveloping function is not a strict power function and decays slower than power law.
At $\alpha=0$, we find $T_{cr} = T_0 \epsilon_{tr}(\Lambda, M, N_b)$, with $T_0/\Delta$ of the order unity (see Fig.4).  

The truncation error of free energy $\delta F = F(M) - F(\infty)$ obeys Eqs.(7) and (8) too, but has two differences. First, the power law extends to $T \ll T_{cr}$, see Fig.1(c). Only very close to $T=0$ is $\delta F$ dominated by a tiny $\delta E_g(M)$ which decreases to zero slower than power law. Second, the quasi-periodic oscillation has a much weaker amplitude than that in $\delta \chi(M)$. Both features facilitate unbiased extrapolation of $F$ to $M=\infty$. Other thermodynamic quantities such as $\langle O \rangle$ and entropy are expected to share these features since they can be obtained from $F$ through derivation of parameters.

The factor $\epsilon_{tr}(\Lambda, M, N_b)$ in the truncation error has a clear physical meaning as the energy scale of the last chain site that can be diagonalized exactly by keeping $M$ states. That this quantity determines the truncation error can be qualitatively understood as this. In the NRG calculation we use a sufficiently long chain with $N \gg N_{T}, N_{M}$.\cite{Weichselbaum2} Here $N_{T} = -\ln{T}/\ln{\Lambda}$ is the chain length corresponding to the temperature $T$ and $N_{M} = \ln{M}/\ln{N_b}$ is the chain length corresponding to $M$. For a fixed $M$, the first $N_{M}$ sites of the chain are diagonalized exactly in NRG calculation. It is the rest $N-N_{M}$ sites in the chain that causes error. The truncation error at high temperature is thus determined by the largest term of the rest chain, i.e., the energy scale $\epsilon_{tr}$. At low temperatures, the truncation error is not as sensitive to $\epsilon_{tr}$ as the high temperature case, because the low energy physics is not determined equally by each site of the chain but mainly by the low energy part. Since the above analysis is based on the universal NRG algorithm, we expect that Eqs.(7) and (8) extracted for SBM applies also to general quantum impurity models.

\begin{figure}[t!]
\vspace{-3.5cm}
\begin{center}
\includegraphics[width=4.8in, height=3.6in, angle=0]{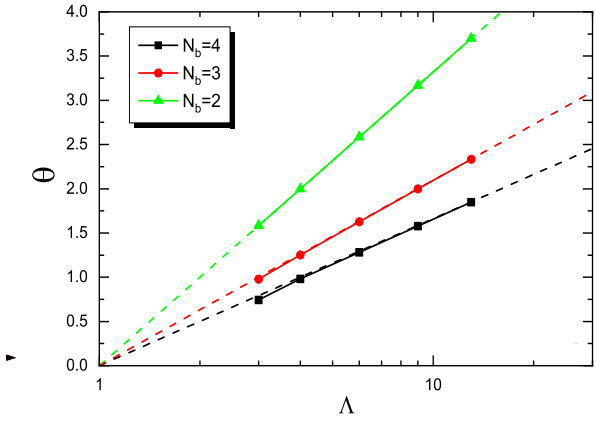}
\vspace*{-1.0cm}
\end{center}
\caption{The power $\theta$ defined by $\delta \chi \propto M^{-\theta}$ as functions of $\Lambda$ for various $N_b$ values. They are fitted from the two largest turning point in the $\chi_{NRG}(M) - \chi_{exact}$ curves, obtained at $\alpha=0.0$ and $T=0.01\Delta \sim 0.1\Delta$ using ENRG. The dashed lines are $y = \ln{x}/\ln{N_b}$.}   \label{Fig3}
\end{figure}

%
\begin{figure}[t!]
\vspace*{-3.5cm}
\begin{center}
\includegraphics[width=5.0in, height=3.8in, angle=0]{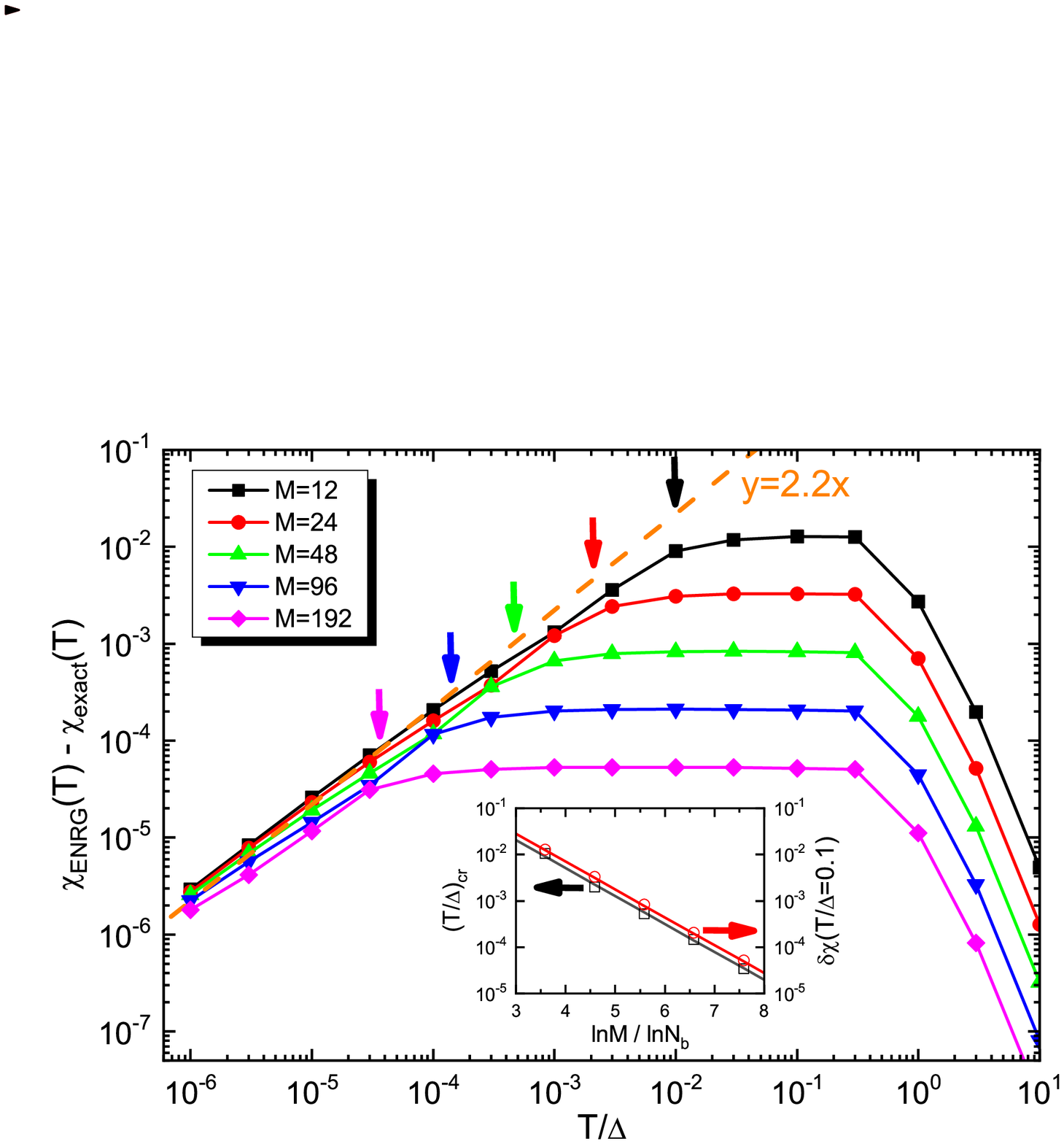}
\vspace*{-1.0cm}
\end{center}
\caption{Truncation error of $\chi$ obtained by ENRG as functions of $T/\Delta$ at $\alpha=0$ for various $M$ values. The arrow of the same color marks out $\left(T/\Delta\right)_{cr}$ for the corresponding $M$. Inset shows $\left(T/\Delta\right)_{cr}$ (black squares, left axis) and $\delta \chi(T/\Delta=0.1)$ (red circles, right axis) as functions of $\ln{M}/\ln{N_b}$. The solid lines are fitting lines of the form $y =c \Lambda^{-x}$. Other parameters are $s=0.3$, $\Delta=0.08$, $\epsilon=0.0$, $\Lambda=4.0$, $N_b=2$, and $N=16$.
}   \label{Fig4}
\end{figure}

In Fig.1, we use the smallest boson states number $N_b=2$ for better illustration. For larger $N_b$, the same rule holds but is more difficult to observe. In Fig.2, we show the truncation error $\delta \chi$ obtained from ENRG at $\alpha=0.0$ and $N_b=4$, for different $\Lambda$ values. Fig.2(a) shows the raw data and Fig.2(b) shows the rescaled ones, clearly illustrating both the period of $\ln{M}/\ln{N_b}$ and the 
relation $\delta \chi \propto \Lambda^{- \ln{M}/\ln{N_b}}$. Note that this dependence can be rewritten into the power law relation $\delta \chi \propto M^{-\ln{\Lambda}/\ln{N_b}}$. One can also see from Fig.2(b) that the periodicity only becomes perfect in the large $M$ limit and that the value of $\Lambda$ influences the phase of the oscillation.

In Fig.3, we show the numerical fitting results for the exponent $\theta$ defined by $\delta \chi \propto M^{-\theta}$. It is obtained from $\alpha=0.0$ at relatively high temperatures. The dependence of $\theta$ on $\Lambda$ and $N_b$ are studied. The good agreement between the data and fitting lines supports the conclusion $\theta = \ln{\Lambda}/\ln{N_b}$, i.e, $\delta \chi \propto M^{-\ln{\Lambda}/\ln{N_b}}$. The less perfect agreement between the data point and the fitting line at $N_b=4$ and $\Lambda=3.0$ is due to the fact that the used temperature $T=0.1\Delta$ is already outside the high temperature regime for this parameter.

\begin{figure}[t!]
\vspace*{-4.0cm}
\begin{center}
\includegraphics[width=5.3in, height=3.8in, angle=0]{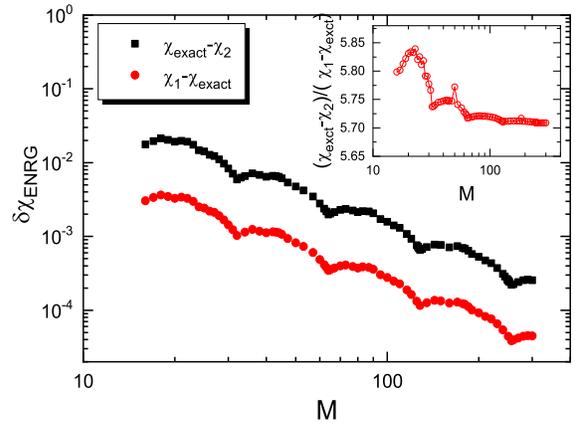}
\vspace*{-1.0cm}
\end{center}
\caption{ Truncation errors $\delta \chi$ for $\chi$'s obtained by two different methods from ENRG as functions of $M$ at $\alpha=0.0$. $\chi_1$ is obtained from the Lehmann representation for Eq.(3) and $\chi_2$ from numerical differential of free energy for Eq.(9). The inset shows their ratio as a function of $M$. Other parameters are $s=0.3$, $\Delta=0.08$, $\epsilon=0.0$, $T=\Delta$, $\Lambda=3.0$, $N_b=2$, and $N=16$. For $\chi_2$, the second-order differential of $F$ is done with $\delta \epsilon=10^{-5}$ and quartic precision calculation.  
}   \label{Fig5}
\end{figure}

To extract the crossover temperature $T_{cr}$, Fig.4 shows the temperature and $M$ dependence of the truncation error $\delta \chi$ at $\alpha=0.0$. For each $M$, we mark out the crossover scale $(T/\Delta)_{cr}$ with an arrow of same color. In the regime $T/\Delta > (T/\Delta)_{cr}$ which includes the intermediate plateau and the large $T/\Delta$ regime, $\delta \chi$ scales as $\epsilon_{tr}=\Lambda^{-\ln{M}/\ln{N_b}}$ for a fixed $T/\Delta$. See, for example, the red circles for $T/\Delta = 0.1$ in the inset. In the regime $T/\Delta < (T/\Delta)_{cr}$, $\delta\chi$ decreases much slower with increasing $M$. $(T/\Delta)_{cr}$ decreases with increasing $M$ in the same power law as $\delta \chi$ does (black squares in the inset). In the inset, both the red and black eye-guiding lines are of the form $y= c\Lambda^{-x}$. The result of Fig.4 is thus summarized as $\delta \chi \propto \epsilon_{tr}$ for $T > T_{cr}$ and $T_{cr} \propto \epsilon_{tr}$. Note that $\delta \chi$ is a linear function of $T/\Delta$ (dashed line in the main figure) in the regime $T/\Delta < (T/\Delta)_{cr}$, which is consistent with the conclusion above.

In Fig.5, the truncation errors of local magnetic susceptibility calculated by two different methods, $\delta \chi_1 = \chi_1(M) - \chi(\infty)$ and $\delta \chi_2 = \chi(\infty) - \chi_2(M)$ are shown as functions of $M$ for $\alpha=0.0$. 
$\chi_1$ is obtained from the ENRG formula for the retarded GF Eq.(3) and $\chi_2$ is from 
\begin{equation}
 \chi = - \partial^2 F / \partial \epsilon^2 |_{\epsilon=0}
\end{equation}
via second-order numerical differential of the ENRG data for $F$. Note $\chi_1(M) \neq \chi_2(M)$ for $M < \infty$ because $\tilde{H}_{N}$ is non-linear in the bias field $\epsilon$ due to the existence of projectors $P_n$. 
But they should approach a common $\chi(\infty)$ in the large $M$ limit.
Fig.5 clearly demonstrates that both errors have the properties summarized in Eqs.(7) and (8), i.e., being quasi-periodic with period $\ln{M}/\ln{N_b}$ and having a decay factor $\epsilon_{tr}(\Lambda, M, N_b)$. They approach $\chi(\infty)$ from opposite directions, i.e., $\chi_1(M) > \chi(\infty)$ while $\chi_2 < \chi(\infty)$. The ratio of $\delta \chi_2$ and $\delta \chi_1$, shown in the inset as a function of $M$, tends to a constant of about $5.7$ in the large $M$ limit. The decaying factor and much of the oscillations in $\delta\chi_1(M)$ and $\delta\chi_2(M)$ cancel but a residual oscillation can still be observed, with decaying amplitude in the large $M$ limit. As to be shown below, this behavior is also present for finite $\alpha$, with much smaller oscillation amplitude.

\section{Truncation Error at $\alpha > 0$} 
 
For finite $\alpha$ values, Fig.6 shows that both $\delta \chi(M)$ and $\delta F(M)$ are proportional to $M^{-\ln{\Lambda}/\ln{N_b}}$ for a high temperature $T \gg T_{cr}$, being consistent with Eqs.(7) and (8). Here, $\chi(\infty)$ and $F(\infty)$ are obtained by fitting the $\delta\chi_2(M)/\delta\chi_1(M)$ data with a constant in the large $M$ limit. In the inset, $|\delta \chi_2(M) / \delta \chi_1(M)|$ approaches a constant value quite fast with increasing $M$, showing the exact cancellation of the $\epsilon_{tr}$ factor and much of the oscillations. The residual oscillation with $M$ in the ratio is much smaller than the case of $\alpha=0$. This provides a fitting method that is more accurate than Eqs.(7) and (8) to obtain $\chi(\infty)$ for finite $\alpha$, i.e., fitting $|\left[\chi_{2}(M)-\chi(\infty) \right]/ \left[\chi(\infty)-\chi_1(M) \right]|$ with a constant in the large $M$ limit. In Fig.6, $\delta F(M)$ has a much weaker oscillation than $\delta \chi(M)$ does. $F(\infty)$ can thus be obtained with high precision by fitting $\delta F(M)$ with Eqs.(7) and (8).

The crossover temperature $T_{cr}$ needs a careful analysis for finite $\alpha$. As known from previous NRG studies, SBM with $0 \leq s \leq 1$ has a localized-delocalized quantum phase transition at $\alpha = \alpha_c$.\cite{Bulla2} On approaching the quantum phase transition, the energy scale of excitations decreases to zero and this must influence the behavior of the truncation error.\cite{Note2} Indeed, for finite $\alpha$, we find that besides a factor $\epsilon_{tr}$, $T_{cr}$ is also proportional to $T^{\ast}$, the crossover energy scale between the stable delocalized fixed point of $\alpha \ll \alpha_c$ and the unstable critical fixed point of $\alpha =\alpha_c$.\cite{Bulla2} $T_{cr}$ is thus summarized for general $\alpha$ as $T_{cr} = T^{\ast} \epsilon_{tr}$. At $\alpha=0$, $T^{\ast}(\alpha)$ reduces to $T_0$. This shows that as $\alpha$ increases towards $\alpha_c$, the decrease of excitation energy scale lifts up the effective temperature, leading to the corresponding decrease of $T_{cr}$. 

This conclusion is supported by Fig.7 and Fig.8.
In Fig.7, the truncation error $\delta \chi$ (calculated by Eq.3) as functions of $\ln{M}/\ln{N_b}$ is shown for $\alpha=0.1$ and a series of temperature. At high temperature $T > T_{cr}$, $\delta \chi \propto \Lambda^{-\ln{M}/\ln{N_b}}$ is verified by the good fitting of line to data (e.g., data for $T/\Delta=1.0$). At low temperatures, $\delta \chi$ does not follow this rule and we find that a crossover $T_{cr}$ is indeed present. To study how $T_{cr}$ depends on NRG parameters for finite $\alpha$, we show in Fig.8 the temperature dependence of $\delta \chi$ for different $\alpha$ values and a fixed $M$ (Fig.8(b)), together with the magnetic susceptibility curves $\chi(T)$ (Fig.8(a)). In Fig.8(b), a peak separates the high temperature regime where $\delta \chi \propto \Lambda^{-\ln{M}/\ln{N_b}}$ from the low temperature regime without this rule. Similar to the curve at $\alpha=0.0$, the $\delta \chi(T)$ curve for finite $\alpha$ is also linear in $T$ in the low temperature regime, supporting the validity of $T_{cr} \propto \epsilon_{tr}$ at finite $\alpha$. Comparison to Fig.8(a) shows that the peak position of $\delta \chi(T)$ in Fig.8(b) corresponds to the saturation temperature of Fig.8(a) below which $\chi(T)$ no longer increases. This temperature scale is just the crossover energy scale $T^{\ast}$ between the unstable critical fixed point at high $T$ and the delocalzied fixed point at low $T$, showing that for finite $\alpha$, $T_{cr} \propto T^{\ast}$. Combining the information, we obtain the final conclusion that $T_{cr} \propto T^{\ast} \epsilon_{tr}$ for $\alpha >0$. This relation naturally recovers that of $\alpha=0$ if we assign $T^{\ast}(\alpha=0)=T_0$.

\begin{figure}[t!]
\vspace{-3.0cm}
\begin{center}
\includegraphics[width=5.0in, height=3.6in, angle=0]{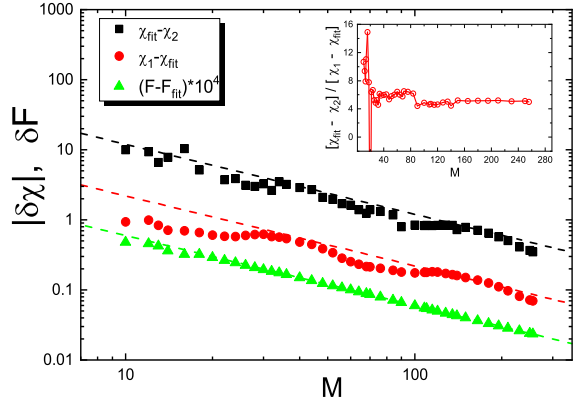}
\vspace*{-1.0cm}
\end{center}
\caption{
The truncation errors $\delta \chi_1(M)$, $\delta \chi_2(M)$, and $\delta F(M)$ from ENRG at $\alpha=0.15$. The dashed lines are eye-guiding curves $y=cx^{-\ln{\Lambda}/\ln{N_b}}$ with different $c$ values. Inset: $|\delta \chi_2 / \delta \chi_1|$ as a function of $M$. For $\delta \chi_2$, we use $\delta \epsilon = 10^{-5}$ and quartic precision calculation. Data are obtained at $s=0.7$, $\Delta=0.01$, $\epsilon=0.0$, $T=0.1\Delta$, $\Lambda=4.0$, $N_b=4$, and $N=28$. The fitted values are $\chi_{\text{fit}}=207.215$ and $F_{\text{fit}}=-0.12773902$.  }  \label{Fig6}
\end{figure}

%
\begin{figure}[t!]
\vspace*{-4.5cm}
\begin{center}
\includegraphics[width=6.0in, height=4.0in, angle=0]{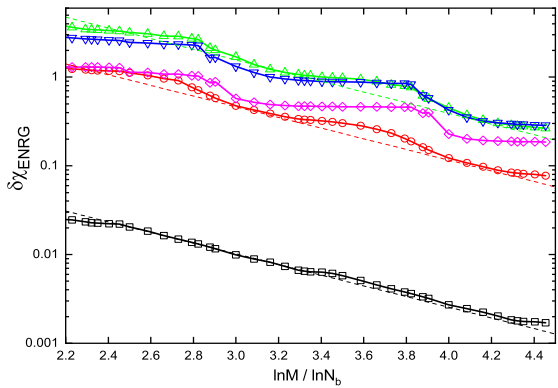}
\vspace*{-1.0cm}
\end{center}
\caption{ Truncation error $\delta \chi = \chi(M) - \chi(\infty)$ obtained from Eq.(3) using ENRG, for $\alpha=0.1$ and temperatures $T/\Delta = 1.0$ (squares), $10^{-1}$ (circles), $10^{-2}$ (up triangles), $10^{-3}$ (down triangles), and $10^{-4}$ (diamonds). The dashed lines are $y= c \Lambda^{- x}$, with $c=0.65$, $29.7$, and $100.0$ from bottom to top. Other parameters are $s=0.7$, $\Delta=0.01$, $\epsilon=0.0$, $\alpha=0.1$, $\Lambda=4.0$, and $N_b=4$. $\chi(\infty)$ is obtained by fitting $|(\chi(\infty)-\chi_2(M))/ (\chi_1(M) - \chi(\infty))|$ with a constant in the large $M$ regime. 
}   \label{Fig7}
\end{figure}

\begin{figure}[t!]
\vspace*{-4.0cm}
\begin{center}
\includegraphics[width=5.7in, height=4.2in, angle=0]{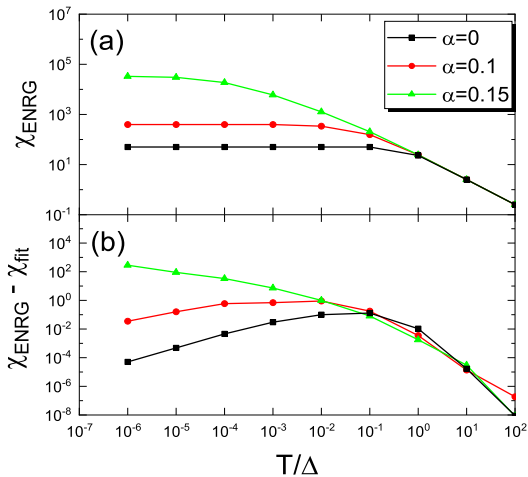}
\vspace*{-1.0cm}
\end{center}
\caption{(a) $\chi$ obtained from ENRG as functions of $T/\Delta$ for various $\alpha$ values. (b) The corresponding truncation error curves as functions of $T/\Delta$. Here the $\chi(\infty)$ values are obtained using the same method as in Fig.7. Other parameters are $s=0.7$, $\Delta=0.01$, $\epsilon=0.0$, $\Lambda=4.0$, $M=216$, and $N_b=4$.
}   \label{Fig8}
\end{figure}
\begin{figure}[t!]
\vspace*{-3.8cm}
\begin{center}
\includegraphics[width=5.9in, height=4.5in, angle=0]{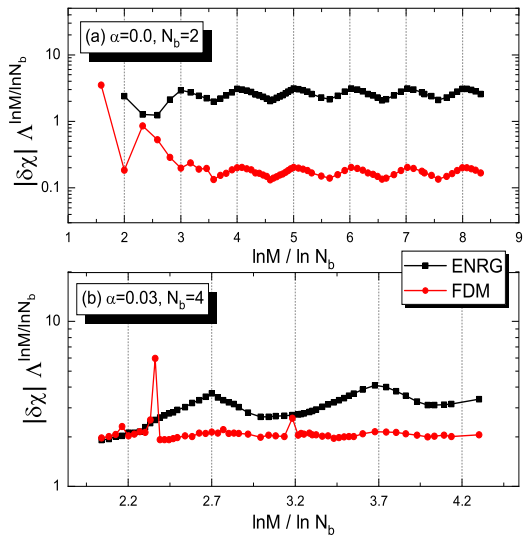}
\vspace*{-1.0cm}
\end{center}
\caption{ Comparison of $\delta \chi \Lambda^{\ln{M}/\ln{N_b}}$ as a function of $\ln{M}/\ln{N_b}$ obtained from ENRG (black squares) and FDM (red circles) methods. (a) is for $\alpha=0.0$ and $N_b=2$; (b) is for $\alpha=0.03$ and $N_b=4$. Other parameters are $s=0.3$, $\Delta=0.08$, $\epsilon=0.0$, $T/\Delta=0.1953125$, $\Lambda=4.0$, and $N=16$. In (b), $\chi(\infty)=8.689$ is obtained by fitting $|(\chi(\infty)-\chi_2(M))/ (\chi_1(M) - \chi(\infty))|$ with a constant in the large $M$ regime. 
}   \label{Fig9}
\end{figure}

Eqs.(7) and (8) are rooted in the structure of $\tilde{H}_{N}(\Lambda, M, N_b)$ and should hold also for other approximate methods for computing physical quantities from the NRG eigen energies and eigen states, if these methods are reasonably accurate. 
In Fig.9, we compare $\delta \chi$'s obtained by Eq.(3) from ENRG and FDM methods. They are calculated at a high temperature $T > T_{cr}$ and for $\alpha=0.0$ (Fig.9(a)) and $\alpha =0.03$ (Fig.9(b)). We find that ENRG produces $\chi(M) > \chi(\infty)$ and FDM NRG produces $\chi(M) < \chi(\infty)$. Both truncation errors are proportional to $\epsilon_{tr}$, with a smaller coefficient for FDM results and better systematics for ENRG at finite $\alpha$. 
More detailed study (now shown) supports the conclusion that $\delta \chi$ from FDM also obeys Eqs.(7) and (8), with only quantitative difference from the ENRG result in the factor $f(\Lambda, x, N_b)$. This is an useful observation since the FDM computation time for $\chi$ scales as $M^{2}N_{b}^{2}N^{1/2}$, much smaller than the scaling $M^{4}N_{b}^{2}N^{2}$ for ENRG.\cite{Yang1} 
We have also checked the data of $\delta \chi$ from the patching method. In contrast to the FDM result, it has poor systematics with $M$ and Eqs.(7) and (8) can not be established for $\delta \chi$ from the patching method.

\section{Discussion and Summary}

In many studies it is necessary to extrapolate the NRG results to the limits $\Lambda=1$ or/and $N_b=\infty$.\cite{Bulla2,Tong1} This is a very difficult task due to the fact that for a fixed $M$, the truncation error increases sharply with decreasing $\Lambda$ or/and with increasing $N_b$. In such situations, the result of present work reminds us that it may be advantageous to carry out extrapolation of $\Lambda$ or/and $N_b$ at constant level of truncation error, i.e., keeping $\epsilon_{tr}(\Lambda, M, N_{b}$ instead of $M$ fixed. This could help separate the effects of truncation error, discretization error, and the boson state truncation error to facilitate the analysis of data. 

Our conclusion Eqs.(7) and (8) for both $\delta \chi$ and $\delta F$ comes from the structure of $\tilde{H}_{N}(\Lambda, M, N_b)$ and has a clear physical picture consistent with the NRG algorithm. Therefore, we expect that the conclusion applies to the NRG study of general quantum impurity models. Note that for other impurity models, one needs to replace the parameter $N_b$ for the SBM with the corresponding Hilbert space dimension of the bath site for the studied model. 

We have analysed the truncation error in $\chi$ and $F$ as functions of NRG parameters. Both of them are static quantities. It is an open question how the truncation error in the dynamical quantities can be analysed quantitatively. Additional complexity arises in such analysis due to the logarithmic broadening conventionally used in the calculation.\cite{Zitko1,Weichselbaum2,Florens1,Zitko3} Considering that $\chi$ is the zero frequency value of the dynamical susceptibility, we expect that such analysis is possible at least for the real part of GF. Further study in this direction is in progress.

In summary, we have analysed the dependence of truncation errors of static physical quantities $\chi$ and $F$ on NRG parameters $M$, $\Lambda$, and $N_b$ for SBM. Quantitative expressions for the truncation errors are extracted for high temperatures, making it possible to accurately extrapolate them to $M=\infty$. Our results are universal for both ENRG and FDM methods and are expected to be applicable to other quantum impurity models.

\section{Acknowledgments}
This work is supported by 973 Program of China (2012CB921704), NSFC grants (11374362, 11974420), Fundamental Research Funds for the Central Universities, and the Research Funds of Renmin University of China 15XNLQ03.



\end{document}